\documentclass[journal=jpclcd,manuscript=letter,10pt]{achemso} 
\usepackage{extsizes}
\mciteErrorOnUnknownfalse
\usepackage{lmodern,mathpazo,amsmath}
\author{Gautier Meyer}
\affiliation{Universit\'e Grenoble Alpes, CNRS, LIPhy, 38000 Grenoble, France}
\author{Ralf Schweins}
\affiliation{Institut Laue Langevin, 38042 Grenoble, France}
\author{Tristan Youngs}
\affiliation{ISIS Rutheford Laboratory, UK}
\author{Jean-Fran\c {c}ois Dufr\^eche}
\affiliation{Institut de Chimie S\'eparative de Marcoules, France}
\author{Isabelle Billard}
\email{isabelle.billard@grenoble-inp.fr}
\affiliation{Universit\'e Grenoble Alpes, Universit\'e Savoie Mont Blanc, CNRS, Grenoble INP, LEPMI, Grenoble, France}
\author{Marie Plazanet}
\email{marie.plazanet@univ-grenoble-alpes.fr}
\affiliation{Univ. Grenoble Alpes, CNRS, LIPhy, 38000 Grenoble, France}

\title{How temperature rise induces phase separation in acidic aqueous biphasic solutions}

\begin{document}

\begin{tocentry}
\includegraphics [width=5cm]{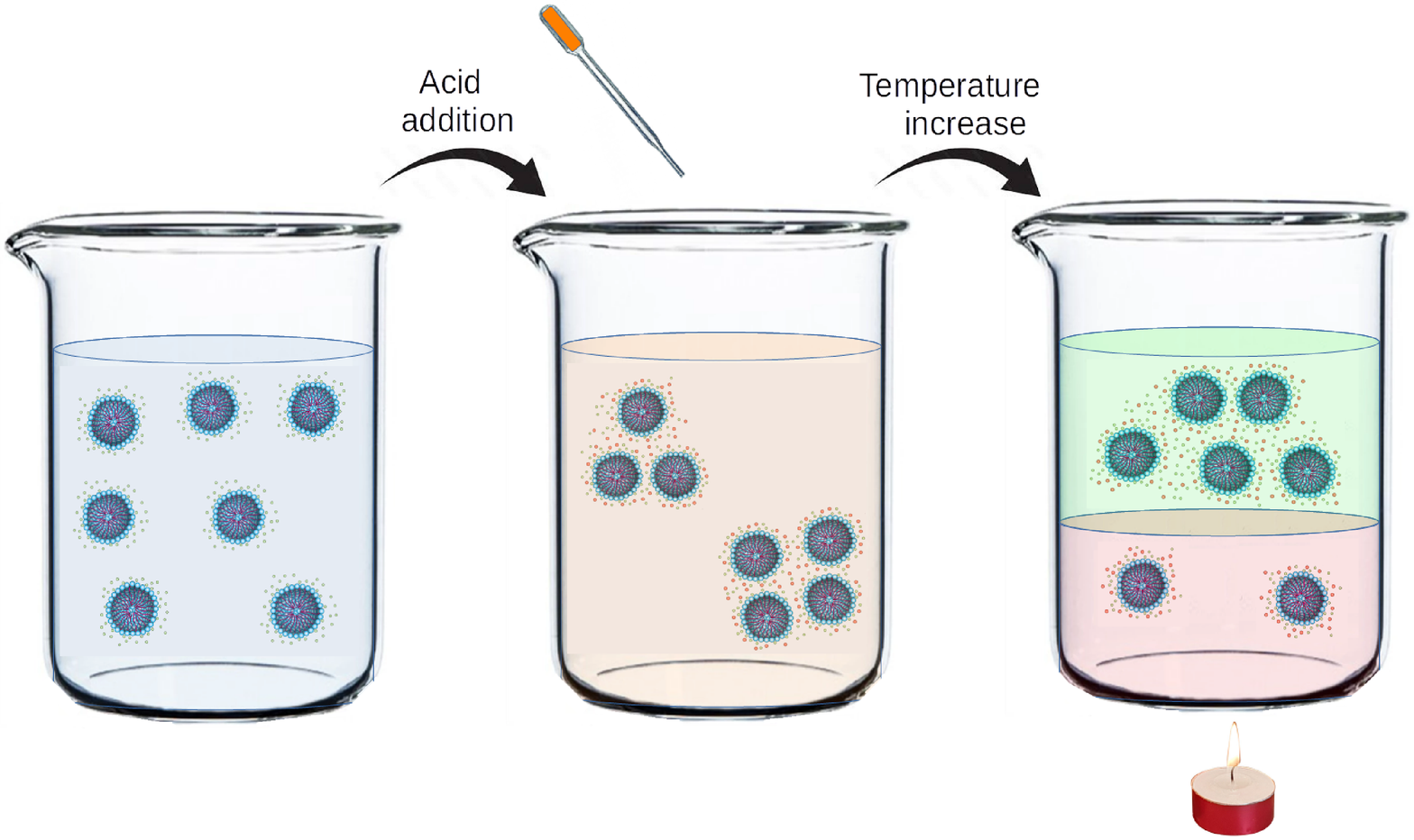}
\end{tocentry}

\begin{abstract}
Ionic-liquid based acidic aqueous biphasic solutions (AcABS) recently offered a breakthrough in the field of metal recycling. Indeed, the mixture of tributyltetradecylphosphonium chloride (P$_{44414}$Cl) and acid with water content larger than 60 \% presents a phase separation with very good extraction efficiency for metallic ions. Moreover, this ternary solution presents a Lower Solution Critical Temperature (LCST), meaning that the biphasic area of the phase diagram increases upon increase of temperature, in other terms the phase separation from a homogeneous liquid can be induced by an elevation of temperature, typically a few tens of degrees. We address here the microscopic mechanisms driving the phase separation. Small Angle Neutron Scattering provides us with structural information for various acid content and temperature. We characterized the spherical micelle formation in the binary ionic liquid/water solution and the micelle aggregation upon addition of acid, due of the screening of electrostatic repulsion. If addition of salt leads to identical transitions in the solution, the ionic strength is not a relevant parameter and more subtle effects such as ion size or polarizability have to be taken into account to rationalize the phase diagram. The increase of both acid concentration and/or temperature eventually leads to the micelle flocculation and phase separation. This last step is achieved through chloride ion adsorption at the surface of the micelle with an enthalpy of adsorption of $\sim$ 12 kJ/mol. The attraction between micelles can be well understood in terms of DLVO potential. This exothermic adsorption compensates the entropic cost, leading to the counter-intuitive behavior of the system. 

\end{abstract}

\section{}
Recycling is one of the biggest tasks of our present society and a very challenging problem. In particular, metallic wastes evolve at a very high rate, due to the exponential growth of technological objects including all kinds of metals, the mixing of which is completely different from what is traditionally found in ores. For this purpose, liquid-liquid extraction (LLE) is an efficient and major way to separate components. The principle is that each chemical, when mixed in the solution, is solvated in a preferential phase of the well chosen two-phase mixture. Recycling should however be performed without the use of polluting solvents, driving the search for green chemistry in this area \citep{Ventura2017,Duan2016,Hilali2019}. In this context, a recent breakthrough was achieved in using aqueous biphasic solutions \citep{metalic_extraction}. Composed by more than ca. 60\% water, ionic liquid (IL) and acid, such a solution (further denoted by AcABS for acidic aqueous biphasic system) avoids the use of carcinogenic, mutagenic and reprotoxic (CMR) solvent while enabling to have large quantities of metallic ions in solution without hydrolysis (depending on ions, for ex. up to 40 g/L for Fe(III) in HCl/H$_2$O solutions \citep{paper_Eris}). Eventually, according to their complexation state, metallic species migrate toward their preferred phase, enabling a very efficient partition \citep{metalic_extraction, Schaeffer2021}.

Such AcABS are thermoresponsive systems. Among them, the mixture of water, tributyltetradecylphosphonium chloride (P$_{44414}$Cl) and acid, also presents the uncommon characteristic of a Lower Critical Solution Temperature (LCST), expressing the property of the solution to separate into two immiscible aqueous phases upon heating. This behavior is opposite to the usual demixion mechanism for which phase separation disappears at high temperature because of the predominance of the entropy of mixing. A more subtle mechanism underneath the phase separation therefore needs to be addressed in order to fully exploit such a phenomenon for extraction purposes, such as metal recycling.
\begin{center}
\begin{figure}
\centering
 \includegraphics[scale=0.5]{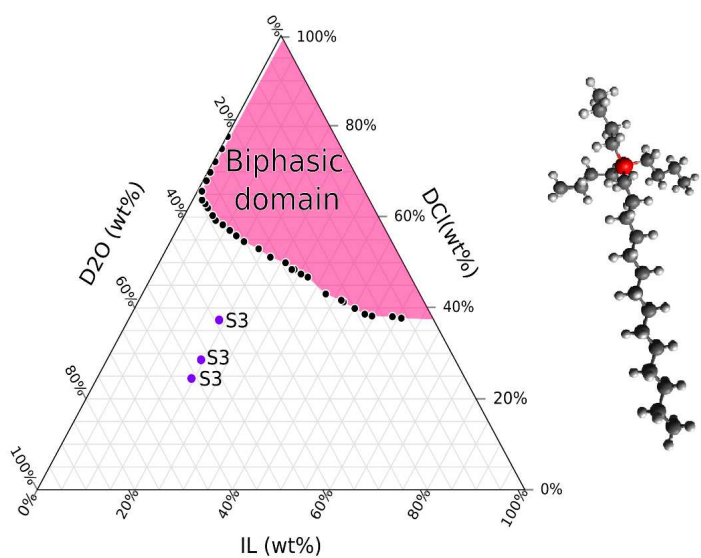}
  \caption{DCl based samples representation on the ternary phase diagram of DCl/P$_{44414}$Cl/D$_2$O ABS at 25$^o$C and schematic representation of the cation P$_{44414}$Cl of the ionic liquid. The structural organization of the three solutions defined by the points S$_1$, S$_2$ and S$_3$ has been studied by SANS. }
  \label{fig:figure1}
\end{figure}
\end{center}

In this letter, we present a combination of structural investigation at the microscopic scale and chloride ion titration, enabling to draw a complete picture of the mechanisms driving the phase separation upon heating. Our system is illustrated by the figure \ref{fig:figure1}. The biphasic region increases with temperature. Interestingly, the system also forms a thermoresponsive ABS in the presence of salt instead of acid, as for example NaCl substituting HCl. The phase diagram presents moreover the same characteristics in terms of \textit{inverse} thermal response in the presence of acid or salt (for ex. HCl or NaCl). Since all the macroscopic observations are similar, the phase transition in presence of acid or salt is expected to be based on the same mechanisms. We therefore used one system or the other depending on the technical requirements of the experiment performed. Moreover, within the precision of our measurements, the binodal is not affected by deuteration of water or acid, also enabling the use of partial deuteration, a great advantage for small angle neutron scattering experiments. Eventually, the nature of the acid has a large influence. In presence of sulfuric acid H$_2$SO$_4$, the thermal response is similar although the biphasic region is increased. The effect is even more striking if nitric acid HNO$_3$ is used, since a single drop turns the IL/water solution in a biphasic one \citep{Vijetha2018}. Such a drastic effect makes however difficult the characterization of the phase transition in presence of HNO$_3$ and will not be discussed in this work.

P$_{44414}$Cl is a compound close to a classical ionic surfactant, for which self assembling in water is expected, similarly to many ionic liquids \citep{Triolo2007,self-assembly,SANS_spheres}. Such an organization of charged units is expected to be influenced by the ionic strength of the solution, modified by the presence of acid, and consequently in the different regions of the phase diagram, since acid concentration also varies. Characterizing the structural organizations of the IL in the different phases should therefore lead us to the microscopic mechanisms of the phase separation. Small angle neutron scattering (SANS) provides a unique tool to probe such a structural organization in solution. The contrast, in our systems, is given by the difference in scattering length between the hydrogenated IL and deuterated water and acid. We will be able to extract both the form of the objects adopted by the self-organized ionic liquids (form factor) simultaneously with the interactions between these objects leading to the structure factor.

\begin{center}
\begin{figure}
\centering
  \includegraphics[scale=0.2]{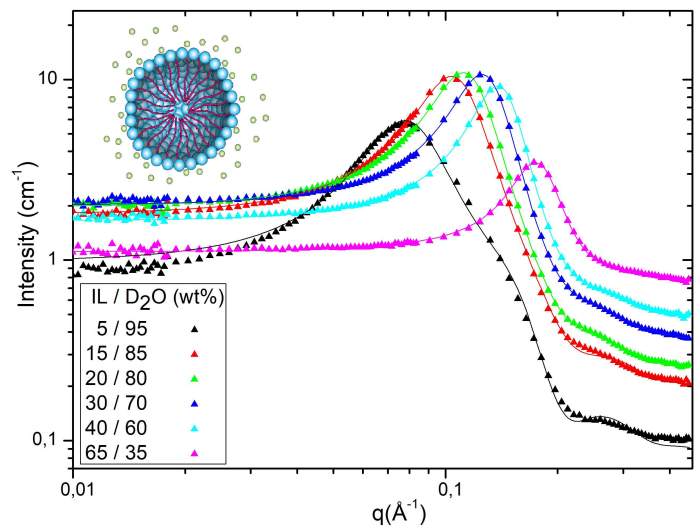}
  \caption{Normalized SANS data (symbols) and fits (solid lines) using a spherical form/hard-sphere structure factors for IL/water mixtures. The insert is a schematic representation of the micelle surrounded by chloride ions.}
  \label{fig:IL_water_raw_fit}
\end{figure}
\end{center}
The first step into the structural investigations was performed on binary IL/water mixtures for various IL contents. The experimental data together with their fitted curves, at 25$^o$C, are shown in figure \ref{fig:IL_water_raw_fit}. The best fit is obtained for a model of spherical micelles formed by the IL, dispersed in suspension because of the repulsive electrostatic interaction modeled by a hard sphere form factor calculated with effective radius (see SI for details). For rather dilute conditions (IL \%wt $<$ 20), the micelle radius has little variations around 19 $\text{\AA{}}$, i.e. roughly the length of a single cation. Considering the charged head and hydrophobic tails of the IL cation, we reasonably assume that the heads of the cations form the external layer of the micelles, surrounded by water and counter ions (Cl$^-$), as schematically represented in the insert of figure \ref{fig:IL_water_raw_fit}. Moreover, charge compensation of the outer surface of the micelle by the chloride anions (counter ions of the IL, unique anions of the solution) is poor since few ions are present to participate to the electrical double layer (EDL) around the micelle. This leads to a residual electrostatic repulsion between the micelles, giving rise to the correlation peak around 0.1 \AA{}$^{-1}$. Note that the peak consistently moves toward higher values of Q, i.e smaller distances, when the IL concentration increases.

Beyond a transition region, for IL content larger than 65\%wt., the microemulsion is inverted and water droplets of radius $\sim$ 5 $\text{\AA{}}$ are surrounded by IL (see figure \ref{fig:micelle-radius} in the SI).

Eventually, measurements were also performed for temperatures increasing up to 55$^o$C. The IL/water solution is homogeneous over this whole temperature range, while it undergoes a phase separation in the presence of acid. No significant changes were indeed observed on the measured intensity as shown figure \ref{fig:IL-Tdep} of the SI. Only a slight increase of S(0) can be extracted and also plotted in figure \ref{fig:osmotic-compressiblity} (SI), which is directly related to the osmotic compressibility \citep{Cousin2015}. Such a trend indicates an increase of the attractive interactions with temperature, however not strong enough to enable the micelle flocculation.

\begin{figure}[H]
  \centering
  \includegraphics[scale=0.15]{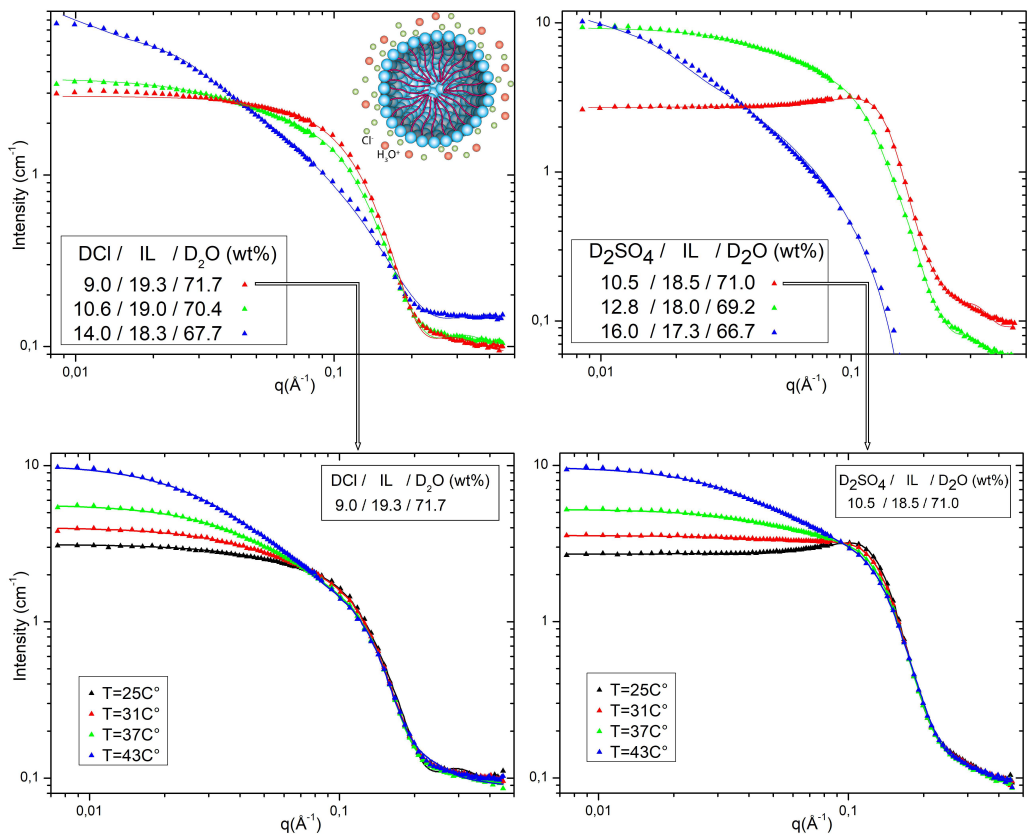}
    \caption{Left panels : measurements on hydrochloric acid solutions; right panels : sulfuric acid solutions. Top : normalized data and fitting results of the SANS measurements performed on ABS solutions at 25$^o$C for three acid contents (S$_1$, S$_2$ and S$_3$ for DCl samples); bottom : data at different temperatures from the monophasic to the biphasic state (upper phase), for transitions at 38 and 36 $^o$C for DCl and D$_2$SO$_4$ solutions respectively at the lowest acid content. The schematic representation of the micelle now contains several types of ions (different colors) in the surrounding electrical double layer.}
      \label{fig:ABS_raw_fit}
\end{figure}

Coming to our system of interest, acid is added to the binary mixture to form the ABS. Data are shown in the figure \ref{fig:ABS_raw_fit} (top panels), for two different acids (hydrochloric acid and sulfuric acid) and three acid contents, at 25$^o$C in monophasic solutions. The weakening of the correlation peak around 0.1 $\text{\AA{}}^{-1}$, stronger when the acid concentration increases, indicates at first sight the screening of the electrostatic interactions between the IL micelles. The effect is similar for both acids. The introduction of other ionic species enables a great decrease of the Debye (screening) length in the EDL. The screening becomes more efficient and the attractive interactions between micelles originating e.g. from Van Der Waals interactions are no longer overwhelmed by the repulsive interactions. However, we emphasize that the structural transformation cannot be rationalized to the ionic strength of the solution. First, the ionic strength is ill-defined in case of sulfuric acid because of its bi-acid nature ; second, if HCl and H$_2$SO$_4$ samples exhibit similar behavior of the structure and transition temperature with acid concentration, it is very different with HNO$_3$ that induces the separation at very low concentration. Indeed, the concept of ionic strength is especially important around the Debye-Huckel regime for which ion interactions are mainly coulombic. In our case, the ion concentration is too high and specific ion interactions play a role. Other effects have to be taken into account such as ion size or polarization or more sophisticated screening approaches due to the high concentration of charges in the solution are necessary \citep{Parsons2011, Sanchez-Fernandez2021}.

\begin{figure}
    \centering
    \includegraphics[scale=0.5]{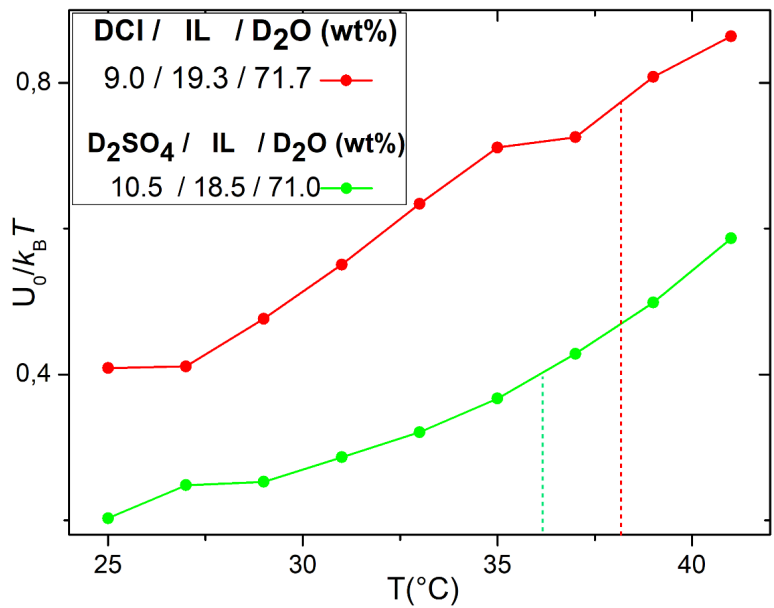}
    \caption{Attractive potential intensity (in k$_B$T) as a function of T for the lower concentration samples. Phase separation temperature is represented by the dashed line of the corresponding color.}
    \label{fig:stickiness_potential}
\end{figure}

Technically, a unique model for fitting the whole set of data, including the high acid contents, could not be found. This strongly suggests that different micellar structures (cylinders, double shell, bicontinuous phase or mixtures of different objects) appear in the solution upon increase of the acidity or ionic strength, as observed in other systems \citep{aswal2003,Sanchez-Fernandez2018,DANOV2021561}. We therefore concentrate here on the low acid content as a model system, giving a clear picture of the process that can be, omitting the form factor details, applied to any region of the phase diagram. 
The best fit is still obtained for a hard sphere form factor, but now combined with a sticky hard sphere structure factor in order to model the attraction between the micelles. The effect of temperature is similar to the addition of acid to the solution, the effect being illustrated by the corresponding top and bottom panels of the figure \ref{fig:ABS_raw_fit}. The screening of electrostatic repulsion between micelles increases and so the stickiness of the structure factor. This temperature behavior is represented in figure \ref{fig:stickiness_potential}.

The stickiness of the potential, directly related to the attractive interaction between the micelles, is therefore the relevant parameter for the description of the phase separation. The potential represented by the sticky hardsphere model is akin to the DLVO potential, a sum a of short ranged attractive potential (VDW) and a repulsive one (electrostatic interactions). The resulting (DLVO) potential presents a barrier that depends on acid content and temperature. When the temperature is high enough so that the thermal energy k$_B$T is close to the height of the barrier, the probability for the micelles to cross the energy barrier and be attracted to each other by Van der Waals interactions increases. In other terms, at high temperature, the micelles eventually flocculate and the phase separation is observed. 

The upper phase is formed by the lowest density phase, i.e. the ionic liquid rich phase. Eventually, the effect is confirmed by the trend of the osmotic compressibility S(0) with temperature, with values way above unity, indicating an increased interaction between the particles and increasing even more with temperature (see SI figure \ref{fig:osmotic-compressiblity}).

\begin{figure}
    \centering
    \includegraphics[scale=0.5]{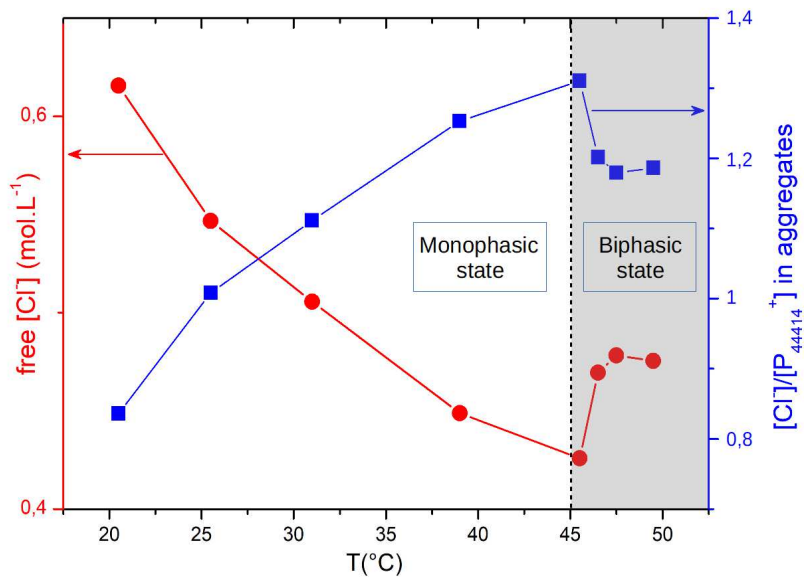}
    \caption{Free [Cl$^-$] in the solution as a function of temperature, assuming that the activity is equal to the concentration (left axis). Corresponding ratio of Cl$^-$ to cations in the EDL of the micelle (right axis) assuming all anions are either free or bound.}
    \label{fig:cloride-concentration}
\end{figure}

Although the screening of electrostatic interactions with the addition of salt or acid is expected, its temperature dependence is less intuitive. The overall increase of the attraction is not believed to be due to stronger Van Der Waals interactions that are poorly temperature dependent and mostly size dependent. It therefore has to be a decrease in the repulsive interaction, meaning a variation in the charge distribution around the micelle or a change of EDL composition. The long distance repulsion between the micelle is typically (in k$B$T units) proportional to the Bjerrum length of the solution. The latter is weakly temperature dependent. Thus the increased attraction at high temperature is not related to dielectric constant changes but rather to the (effective)charge of the micelles, i.e. to the ion adsorption at the surface.
In order to address this last point in solutions where the zeta-potential measurement does not provide any reliable values, we performed a titration of the chloride Cl$^-$ activity in the solution. We used a specific electrode enabling the measurement of Cl- concentration in the solution, directly converted, thanks to an appropriate calibration, in free Cl$^-$ concentration. We consider here the overall measured activity to be equal to the activity of the free chloride, excluding the activity of the chloride that are bound to a micelle surface. The activity of the free Cl$^-$ is moreover considered as being equal to their concentration. 
Because of experimental drawbacks, these measurements were performed using NaCl instead of HCl, since the mechanisms are assumed to be similar in both mixtures. Results are shown in the figure \ref{fig:cloride-concentration}. Increasing the temperature, the concentration of free Cl$^-$ in solution decreases, meaning an increase of [Cl$^-$] in the EDL. Knowing the total concentration of Cl$^-$, we extract the quantitative variation of Cl$^-$ that are bound to a micelle.

The adsorption of Cl$^-$ in the EDL, following the interpretation of the structural data, leads to a decrease in the repulsive interactions, easing micelles flocculation.
Assuming an Arrhenius behavior for this adsorption process and two possibilities only for the Cl$^-$ (free or bound), we get the enthalpy of adsorption of chloride ions on the micelles to be around 12 kJ/mol.

A complete picture of the microscopic mechanisms leading to the phase separation induced by temperature can now be drawn. Most probably because of the increasing disorder of the short aliphatic chains around the phosphonium ions of the ionic liquid, chloride ions are attracted to the surface of the micelle, increasing the screening of electrostatic repulsion that maintains the micelles in suspension. 

Once the effective energy barrier of the DLVO potential is low enough, the micelles are able to reversibly aggregate and form a biphasic system. The phase separation is thus a kind of liquid-gas separation of a fluid of micelles. At high temperature, if the magnitude of the micelle configurational entropy increases, it is no longer sufficient to compensate for the enthalpy gain, which also increases due to the adsorption of ions. This enthalpic gain compensates the entropic cost of the phase separation upon temperature increase.

Such a phase separation upon rise of temperature is unusual but not unique, as found in LCST systems. Although more commonly encountered in polymeric systems \citep{Lee2010,Bruce2019}, other mixtures of small molecules, including ionic liquids exhibit such behavior \citep{Plazanet2004-aCD4MP,Fukumoto2007,Hoogerstraete2013, Wu2020}. Moreover, ionic liquid based solvents are prone to complex behaviors because of the high charge density, leading most usual polyelectrolyte description inappropriate \citep{Sanchez-Fernandez2021}.  
The methodology presented here would therefore apply to many systems of soft matter, providing a bridging description from the nanoscale to macroscopic properties.

\begin{acknowledgement}

We thank the IDEX University Grenoble-Alpes for funding the PhD of G. Meyer (IRS RIMMEL). We also thank the Institut Laue-Langevin (Grenoble, France) for the allocation of neutron beam time. The neutron data set is available at DOI: 10.5291/ILL-DATA.6-02-604. 

\end{acknowledgement}

\bibliography{biblio}

\providecommand{\latin}[1]{#1}
\makeatletter
\providecommand{\doi}
  {\begingroup\let\do\@makeother\dospecials
  \catcode`\{=1 \catcode`\}=2 \doi@aux}
\providecommand{\doi@aux}[1]{\endgroup\texttt{#1}}
\makeatother
\providecommand*\mcitethebibliography{\thebibliography}
\csname @ifundefined\endcsname{endmcitethebibliography}
  {\let\endmcitethebibliography\endthebibliography}{}
\begin{mcitethebibliography}{24}
\providecommand*\natexlab[1]{#1}
\providecommand*\mciteSetBstSublistMode[1]{}
\providecommand*\mciteSetBstMaxWidthForm[2]{}
\providecommand*\mciteBstWouldAddEndPuncttrue
  {\def\EndOfBibitem{\unskip.}}
\providecommand*\mciteBstWouldAddEndPunctfalse
  {\let\EndOfBibitem\relax}
\providecommand*\mciteSetBstMidEndSepPunct[3]{}
\providecommand*\mciteSetBstSublistLabelBeginEnd[3]{}
\providecommand*\EndOfBibitem{}
\mciteSetBstSublistMode{f}
\mciteSetBstMaxWidthForm{subitem}{(\alph{mcitesubitemcount})}
\mciteSetBstSublistLabelBeginEnd
  {\mcitemaxwidthsubitemform\space}
  {\relax}
  {\relax}

\bibitem[Ventura \latin{et~al.}(2017)Ventura, e~Silva, Quental, Mondal, Freire,
  and Coutinho]{Ventura2017}
Ventura,~S. P.~M.; e~Silva,~F.~A.; Quental,~M.~V.; Mondal,~D.; Freire,~M.~G.;
  Coutinho,~J. A.~P. Ionic-Liquid-Mediated Extraction and Separation Processes
  for Bioactive Compounds: Past, Present, and Future Trends. \emph{Chemical
  Reviews} \textbf{2017}, \emph{117}, 6984--7052, PMID: 28151648\relax
\mciteBstWouldAddEndPuncttrue
\mciteSetBstMidEndSepPunct{\mcitedefaultmidpunct}
{\mcitedefaultendpunct}{\mcitedefaultseppunct}\relax
\EndOfBibitem
\bibitem[Duan \latin{et~al.}(2016)Duan, Dou, Guo, Li, and Liu]{Duan2016}
Duan,~L.; Dou,~L.-L.; Guo,~L.; Li,~P.; Liu,~E.-H. Comprehensive Evaluation of
  Deep Eutectic Solvents in Extraction of Bioactive Natural Products. \emph{ACS
  Sustainable Chemistry \& Engineering} \textbf{2016}, \emph{4},
  2405--2411\relax
\mciteBstWouldAddEndPuncttrue
\mciteSetBstMidEndSepPunct{\mcitedefaultmidpunct}
{\mcitedefaultendpunct}{\mcitedefaultseppunct}\relax
\EndOfBibitem
\bibitem[Hilali \latin{et~al.}(2019)Hilali, Fabiano-Tixier, Ruiz, Hejjaj,
  Ait~Nouh, Idlimam, Bily, Mandi, and Chemat]{Hilali2019}
Hilali,~S.; Fabiano-Tixier,~A.-S.; Ruiz,~K.; Hejjaj,~A.; Ait~Nouh,~F.;
  Idlimam,~A.; Bily,~A.; Mandi,~L.; Chemat,~F. Green Extraction of Essential
  Oils, Polyphenols, and Pectins from Orange Peel Employing Solar Energy:
  Toward a Zero-Waste Biorefinery. \emph{ACS Sustainable Chemistry \&
  Engineering} \textbf{2019}, \emph{7}, 11815--11822\relax
\mciteBstWouldAddEndPuncttrue
\mciteSetBstMidEndSepPunct{\mcitedefaultmidpunct}
{\mcitedefaultendpunct}{\mcitedefaultseppunct}\relax
\EndOfBibitem
\bibitem[Gras \latin{et~al.}(2018)Gras, Papaiconomou, Schaeffer, Cha{\^i}net,
  Tedjar, Coutinho, and Billard]{metalic_extraction}
Gras,~M.; Papaiconomou,~N.; Schaeffer,~N.; Cha{\^i}net,~E.; Tedjar,~F.;
  Coutinho,~J.~A.; Billard,~I. {Ionic-Liquid-Based Acidic Aqueous Biphasic
  Systems for Simultaneous Leaching and Extraction of Metallic Ions}.
  \emph{{Angewandte Chemie International Edition}} \textbf{2018}, \emph{57},
  1563 -- 1566\relax
\mciteBstWouldAddEndPuncttrue
\mciteSetBstMidEndSepPunct{\mcitedefaultmidpunct}
{\mcitedefaultendpunct}{\mcitedefaultseppunct}\relax
\EndOfBibitem
\bibitem[Sinoimeri \latin{et~al.}(2020)Sinoimeri, Maia~Fernandes, Cognard,
  Pereira, Svecova, Guillotte, and Billard]{paper_Eris}
Sinoimeri,~E.; Maia~Fernandes,~V.; Cognard,~J.; Pereira,~J. F.~B.; Svecova,~L.;
  Guillotte,~I.; Billard,~I. Uncommon biphasic behaviour induced by very high
  metal ion concentrations in HCl/H2O/[P44414]Cl and HCl/H2O/PEG-600 systems.
  \emph{Phys. Chem. Chem. Phys.} \textbf{2020}, \emph{22}, 23226--23236\relax
\mciteBstWouldAddEndPuncttrue
\mciteSetBstMidEndSepPunct{\mcitedefaultmidpunct}
{\mcitedefaultendpunct}{\mcitedefaultseppunct}\relax
\EndOfBibitem
\bibitem[Schaeffer \latin{et~al.}(2021)Schaeffer, Vargas, Passos, Brandão,
  Nogueira, Svecova, Papaiconomou, and Coutinho]{Schaeffer2021}
Schaeffer,~N.; Vargas,~S. J.~R.; Passos,~H.; Brandão,~P.; Nogueira,~H. I.~S.;
  Svecova,~L.; Papaiconomou,; Coutinho,~J. A.~P. A HNO3-Responsive Aqueous
  Biphasic System for Metal Separation: Application towards CeIV Recovery.
  \emph{ChemSusChem} \textbf{2021}, \emph{14}, 3018--3026\relax
\mciteBstWouldAddEndPuncttrue
\mciteSetBstMidEndSepPunct{\mcitedefaultmidpunct}
{\mcitedefaultendpunct}{\mcitedefaultseppunct}\relax
\EndOfBibitem
\bibitem[Mogilireddy \latin{et~al.}(2018)Mogilireddy, Gras, Schaeffer, Passos,
  Svecova, Papaiconomou, Coutinho, and Billard]{Vijetha2018}
Mogilireddy,~V.; Gras,~M.; Schaeffer,~N.; Passos,~H.; Svecova,~L.;
  Papaiconomou,~N.; Coutinho,~J. A.~P.; Billard,~I. Understanding the
  fundamentals of acid-induced ionic liquid-based aqueous biphasic system.
  \emph{Phys. Chem. Chem. Phys.} \textbf{2018}, \emph{20}, 16477--16484\relax
\mciteBstWouldAddEndPuncttrue
\mciteSetBstMidEndSepPunct{\mcitedefaultmidpunct}
{\mcitedefaultendpunct}{\mcitedefaultseppunct}\relax
\EndOfBibitem
\bibitem[Triolo \latin{et~al.}(2007)Triolo, Russina, Bleif, and
  Di~Cola]{Triolo2007}
Triolo,~A.; Russina,~O.; Bleif,~H.-J.; Di~Cola,~E. Nanoscale Segregation in
  Room Temperature Ionic Liquids. \emph{The Journal of Physical Chemistry B}
  \textbf{2007}, \emph{111}, 4641--4644\relax
\mciteBstWouldAddEndPuncttrue
\mciteSetBstMidEndSepPunct{\mcitedefaultmidpunct}
{\mcitedefaultendpunct}{\mcitedefaultseppunct}\relax
\EndOfBibitem
\bibitem[Greaves and Drummond(2013)Greaves, and Drummond]{self-assembly}
Greaves,~T.~L.; Drummond,~C.~J. Solvent nanostructure{,} the solvophobic effect
  and amphiphile self-assembly in ionic liquids. \emph{Chem. Soc. Rev.}
  \textbf{2013}, \emph{42}, 1096--1120\relax
\mciteBstWouldAddEndPuncttrue
\mciteSetBstMidEndSepPunct{\mcitedefaultmidpunct}
{\mcitedefaultendpunct}{\mcitedefaultseppunct}\relax
\EndOfBibitem
\bibitem[J.~Bowers and Vergara-Gutierrez(2004)J.~Bowers, and
  Vergara-Gutierrez]{SANS_spheres}
J.~Bowers,~P. J.~M.,~C. P.~Butts; Vergara-Gutierrez,~M.~C. Aggregation Behavior
  of Aqueous Solutions of Ionic Liquids. \emph{Langmuir} \textbf{2004},
  \emph{20}, 2191--2198\relax
\mciteBstWouldAddEndPuncttrue
\mciteSetBstMidEndSepPunct{\mcitedefaultmidpunct}
{\mcitedefaultendpunct}{\mcitedefaultseppunct}\relax
\EndOfBibitem
\bibitem[{Cousin, Fabrice}(2015)]{Cousin2015}
{Cousin, Fabrice}, Small angle neutron scattering. \emph{EPJ Web of
  Conferences} \textbf{2015}, \emph{104}, 01004\relax
\mciteBstWouldAddEndPuncttrue
\mciteSetBstMidEndSepPunct{\mcitedefaultmidpunct}
{\mcitedefaultendpunct}{\mcitedefaultseppunct}\relax
\EndOfBibitem
\bibitem[Parsons \latin{et~al.}(2011)Parsons, Boström, Nostro, and
  Ninham]{Parsons2011}
Parsons,~D.~F.; Boström,~M.; Nostro,~P.~L.; Ninham,~B.~W. Hofmeister effects:
  interplay of hydration{,} nonelectrostatic potentials{,} and ion size.
  \emph{Phys. Chem. Chem. Phys.} \textbf{2011}, \emph{13}, 12352--12367\relax
\mciteBstWouldAddEndPuncttrue
\mciteSetBstMidEndSepPunct{\mcitedefaultmidpunct}
{\mcitedefaultendpunct}{\mcitedefaultseppunct}\relax
\EndOfBibitem
\bibitem[Sanchez-Fernandez \latin{et~al.}(2021)Sanchez-Fernandez, Jackson,
  Prévost, Doutch, and Edler]{Sanchez-Fernandez2021}
Sanchez-Fernandez,~A.; Jackson,~A.~J.; Prévost,~S.~F.; Doutch,~J.~J.;
  Edler,~K.~J. Long-Range Electrostatic Colloidal Interactions and Specific Ion
  Effects in Deep Eutectic Solvents. \emph{Journal of the American Chemical
  Society} \textbf{2021}, \emph{143}, 14158--14168, PMID: 34459188\relax
\mciteBstWouldAddEndPuncttrue
\mciteSetBstMidEndSepPunct{\mcitedefaultmidpunct}
{\mcitedefaultendpunct}{\mcitedefaultseppunct}\relax
\EndOfBibitem
\bibitem[Aswal and Goyal(2003)Aswal, and Goyal]{aswal2003}
Aswal,~V.; Goyal,~P. Role of different counterions and size of micelle in
  concentration dependence micellar structure of ionic surfactants.
  \emph{Chemical physics letters} \textbf{2003}, \emph{368}, 59--65\relax
\mciteBstWouldAddEndPuncttrue
\mciteSetBstMidEndSepPunct{\mcitedefaultmidpunct}
{\mcitedefaultendpunct}{\mcitedefaultseppunct}\relax
\EndOfBibitem
\bibitem[Sanchez-Fernandez \latin{et~al.}(2018)Sanchez-Fernandez, Hammond,
  Edler, Arnold, Doutch, Dalgliesh, Li, Ma, and Jackson]{Sanchez-Fernandez2018}
Sanchez-Fernandez,~A.; Hammond,~O.~S.; Edler,~K.~J.; Arnold,~T.; Doutch,~J.;
  Dalgliesh,~R.~M.; Li,~P.; Ma,~K.; Jackson,~A.~J. Counterion binding alters
  surfactant self-assembly in deep eutectic solvents. \emph{Phys. Chem. Chem.
  Phys.} \textbf{2018}, \emph{20}, 13952--13961\relax
\mciteBstWouldAddEndPuncttrue
\mciteSetBstMidEndSepPunct{\mcitedefaultmidpunct}
{\mcitedefaultendpunct}{\mcitedefaultseppunct}\relax
\EndOfBibitem
\bibitem[Danov \latin{et~al.}(2021)Danov, Kralchevsky, Stanimirova, Stoyanov,
  Cook, and Stott]{DANOV2021561}
Danov,~K.~D.; Kralchevsky,~P.~A.; Stanimirova,~R.~D.; Stoyanov,~S.~D.;
  Cook,~J.~L.; Stott,~I.~P. Analytical modeling of micelle growth. 4. Molecular
  thermodynamics of wormlike micelles from ionic surfactants: Theory vs.
  experiment. \emph{Journal of Colloid and Interface Science} \textbf{2021},
  \emph{584}, 561--581\relax
\mciteBstWouldAddEndPuncttrue
\mciteSetBstMidEndSepPunct{\mcitedefaultmidpunct}
{\mcitedefaultendpunct}{\mcitedefaultseppunct}\relax
\EndOfBibitem
\bibitem[Lee and Lodge(2010)Lee, and Lodge]{Lee2010}
Lee,~H.-N.; Lodge,~T.~P. Lower Critical Solution Temperature (LCST) Phase
  Behavior of Poly(ethylene oxide) in Ionic Liquids. \emph{The Journal of
  Physical Chemistry Letters} \textbf{2010}, \emph{1}, 1962--1966\relax
\mciteBstWouldAddEndPuncttrue
\mciteSetBstMidEndSepPunct{\mcitedefaultmidpunct}
{\mcitedefaultendpunct}{\mcitedefaultseppunct}\relax
\EndOfBibitem
\bibitem[Bruce \latin{et~al.}(2019)Bruce, Bui, Rogers, Cremer, and van~der
  Vegt]{Bruce2019}
Bruce,~E.~E.; Bui,~P.~T.; Rogers,~B.~A.; Cremer,~P.~S.; van~der Vegt,~N. F.~A.
  Nonadditive Ion Effects Drive Both Collapse and Swelling of Thermoresponsive
  Polymers in Water. \emph{Journal of the American Chemical Society}
  \textbf{2019}, \emph{141}, 6609--6616, PMID: 30919630\relax
\mciteBstWouldAddEndPuncttrue
\mciteSetBstMidEndSepPunct{\mcitedefaultmidpunct}
{\mcitedefaultendpunct}{\mcitedefaultseppunct}\relax
\EndOfBibitem
\bibitem[{\underline{Plazanet}} \latin{et~al.}(2004){\underline{Plazanet}},
  Floare, Johnson, Schweins, and Trommsdorff]{Plazanet2004-aCD4MP}
{\underline{Plazanet}},~M.; Floare,~C.; Johnson,~M.~R.; Schweins,~R.;
  Trommsdorff,~H.~P. {Freezing on heating of liquid solutions}. \emph{Journal
  of Chemical Physics} \textbf{2004}, \emph{121}, 5031--5034\relax
\mciteBstWouldAddEndPuncttrue
\mciteSetBstMidEndSepPunct{\mcitedefaultmidpunct}
{\mcitedefaultendpunct}{\mcitedefaultseppunct}\relax
\EndOfBibitem
\bibitem[Fukumoto and Ohno(2007)Fukumoto, and Ohno]{Fukumoto2007}
Fukumoto,~K.; Ohno,~H. LCST-Type Phase Changes of a Mixture of Water and Ionic
  Liquids Derived from Amino Acids. \emph{Angewandte Chemie International
  Edition} \textbf{2007}, \emph{46}, 1852--1855\relax
\mciteBstWouldAddEndPuncttrue
\mciteSetBstMidEndSepPunct{\mcitedefaultmidpunct}
{\mcitedefaultendpunct}{\mcitedefaultseppunct}\relax
\EndOfBibitem
\bibitem[Vander~Hoogerstraete \latin{et~al.}(2013)Vander~Hoogerstraete,
  Onghena, and Binnemans]{Hoogerstraete2013}
Vander~Hoogerstraete,~T.; Onghena,~B.; Binnemans,~K. Homogeneous
  Liquid–Liquid Extraction of Metal Ions with a Functionalized Ionic Liquid.
  \emph{The Journal of Physical Chemistry Letters} \textbf{2013}, \emph{4},
  1659--1663, PMID: 26282975\relax
\mciteBstWouldAddEndPuncttrue
\mciteSetBstMidEndSepPunct{\mcitedefaultmidpunct}
{\mcitedefaultendpunct}{\mcitedefaultseppunct}\relax
\EndOfBibitem
\bibitem[Wu \latin{et~al.}(2020)Wu, Zhang, Deng, Li, Luo, Zheng, and
  Dong]{Wu2020}
Wu,~S.; Zhang,~Q.; Deng,~Y.; Li,~X.; Luo,~Z.; Zheng,~B.; Dong,~S. Assembly
  Pattern of Supramolecular Hydrogel Induced by Lower Critical Solution
  Temperature Behavior of Low-Molecular-Weight Gelator. \emph{Journal of the
  American Chemical Society} \textbf{2020}, \emph{142}, 448--455, PMID:
  31825602\relax
\mciteBstWouldAddEndPuncttrue
\mciteSetBstMidEndSepPunct{\mcitedefaultmidpunct}
{\mcitedefaultendpunct}{\mcitedefaultseppunct}\relax
\EndOfBibitem
\bibitem[Guinier and Fournet(1955)Guinier, and Fournet]{intensity_sphere}
Guinier,~A.; Fournet,~G. Small-Angle Scattering of X-Rays. \emph{John Wiley and
  Sons} \textbf{1955}, \relax
\mciteBstWouldAddEndPunctfalse
\mciteSetBstMidEndSepPunct{\mcitedefaultmidpunct}
{}{\mcitedefaultseppunct}\relax
\EndOfBibitem
\end{mcitethebibliography}

\begin{suppinfo}

\subsection{Materials and methods}

\textbf{Samples.}Tributyltetradecylphosphonium chloride was purchased from IoLiTec guaranting a purity higher than 95\%. Several mineral acids were used, in their deuterated form, together with D$_2$O to optimize the contrast in neutron scattering (see methods). Deuterated water (99,90\% deuterium), D$_2$SO$_4$ (98 wt\% in D$_2$O, $>99\%$ deuterium) and and DNO$_3$ (65-70 wt\% in D$_2$O, $>99\%$ deuterium) per purshased from Eurisotop and DCl (35 wt\% in D$_2$O, $>99\%$ deuterated) from Sigma-Aldrich.

The ABS samples phase transition temperatures were determined using the cloud point method, where the monophasic samples temperature is increased until turbidity is observed.

\textbf{Chloride titration.} Chloride concentrations have been measured with a Cl$^-$ -specific electrode (Thermo Scientific, chloride half-cell Orion 9417SC and reference cell). A linear calibration has been obtained with four NaCl aqueous solutions in the range 10$^{-1}$ M / 1 M  at five temperatures in the range 19$^o$C - 54 $^o$C. It was further checked with P$_{4444}$Cl aqueous solutions.

\textbf{SANS methods.} SANS measurements on IL/water solutions have been carried out on the diffractometer NIMROD at the ISIS neutron source (Oxford, UK) and on the small-angle neutron scattering instrument D11 at the Institut Laue-Langevin (ILL, Grenoble, France).

SANS experiments on the D11 instruments were performed under the following conditions. Neutrons have been measured using a 3He MWPC (CERCA) detector with 256 x 256 pixels of 3.75 mm x 3.75 mm size. Three instrument configurations have been employed, all with a neutron wavelength of 6 Å (FWHM 9$\%$), and sample  detector distances of 34 m, 8 m and 1.4 m (with collimation distances of 34 m, 8 m and 20.5 m respectively). A fourth setting has been used for time-resolved experiments, with a sample  detector distance of 4 m and a collimation distance of 20.5 m. The neutron footprint on the sample was 16 mm in diameter.
The cuvettes had an internal size of 45 mm x 45 mm and were made by Thuet (France). All data were put on absolute scale making use of a 1 mm H$_2$O measurement, serving as secondary calibration standard, and having a differential scattering cross section of 0.983 cm$^{-1}$.

Selective deuteration makes SANS a very powerful technique to characterize aggregates in solution through various parameters such as their shape, size or volume fraction. The hydrogen/deuterium scattering length density (SLD) contrast is used to maximize the SLD contrast between the IL cation and the surrounding solvent, reducing simultaneously the background due to the incoherent neutron scattering of hydrogen. In the present case, we used hydrogenated IL and deuterated solvents. The scattered neutron intensity is measured as a function of the wave vector transfer momentum $q=\frac{4\pi \sin(\frac{\theta}{2})}{\lambda}$. The q-range covered respectively on NIMROD and D11 are $0.05 \text{\AA{}}$ - $5 \text{\AA{}}$ and $0.0018 \text{\AA{}}$ - $0.45 \text{\AA{}}$. Such an extended q-range enables to determine molecular structural information on the shorter scale (high q), while contrariwise the supramolecular organization is probed at a larger scale at smaller q. Eventually, the lowest Q-range was measured only for some selected samples and did not provide any additional information and are therefore not presented here.

NIMROD data were reduced with the program GUDRUN. D11 data reduction was performed with the LAMP software (https://www.ill.eu/fr/users/support-labs-infrastructure/software-scientific-tools/lamp), while the dedicated software SASView was used for model fitting, enabling the combination of different models for both structure and form factors. For the various samples and conditions, the best fit to the measured intensity was obtained in combining the hardsphere form factor with a structure factor of the form either hard sphere with electrostatic interactions (meaning a hard sphere structure factor with an effective radius different than the one used for the form factor), or a sticky hard sphere potential. 

The intensity scattered by a spherical object in a solution is given by \citep{intensity_sphere}:

$$I(q)=\frac{V_f}{V}.\left( 3V( \Delta \rho). \frac{\sin(qR)-qR \cos(q{R}))}{(qR)^3}\right)^2+B$$

where $V_f$ is the volume fraction, $V$ is the volume of the scatterer, $R$ is the sphere radius, $B$ is the background level and $\Delta \rho$ is the scattering length density difference between the scatterer and the solvent.\\

Polydispersity of the spheres was accounted for in the modeling process. The distribution of radii is considered to follow a gaussian profile, and the polydispersity $PD$ parameter is given by: $PD=\sigma/\overline{x}$ , where $\sigma$ is the standard deviation of the Gaussian distribution and $\overline{x}$ is its center value.

\subsection{Complement on the SANS analysis of binary IL/water solutions}
The data were fitted using a hard sphere model for the form factor and hard sphere with electrostatic interactions, leading to an effective radius for the structure factor. The relevant parameters, radius (form factor, figure \ref{fig:micelle-radius}) and effective radius (structure factor, figure \ref{fig:micelle-effective-radius}) are plotted in the following figures. The polydispersity is taken into account, as represented in the figure \ref{fig:micelle-polydispersity}.

\begin{figure}
    \centering
    \includegraphics[scale=0.3]{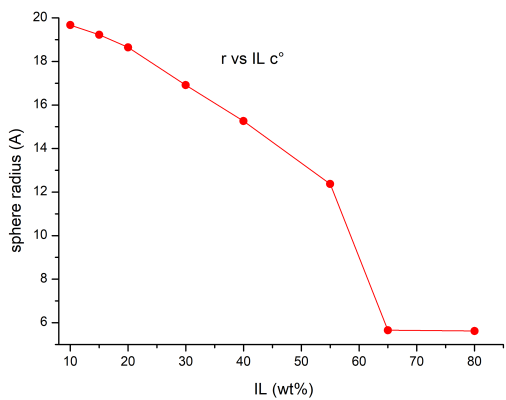}
    \caption{IL/water : radius of the micelle form factor vs IL content}
    \label{fig:micelle-radius}
\end{figure}

\begin{figure}
    \centering
    \includegraphics[scale=0.3]{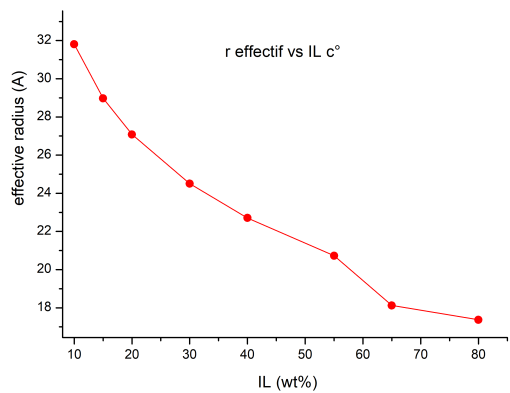}
    \caption{IL/water : effective radius of the hard-sphere structure factor vs IL content}
    \label{fig:micelle-effective-radius}
\end{figure}

\begin{figure}
    \centering
    \includegraphics[scale=0.3]{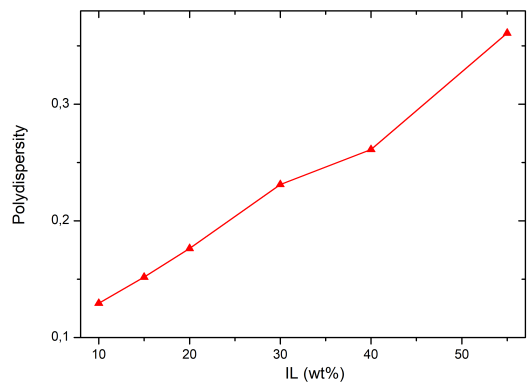}
    \caption{IL/water : polydispersity of the radius of the micelle as a function of IL content}
    \label{fig:micelle-polydispersity}
\end{figure}

\begin{figure}
    \centering
    \includegraphics[scale=0.3]{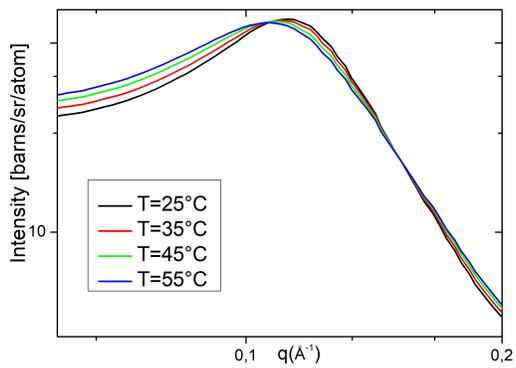}
    \caption{Temperature dependence of SANS data of the IL/water 20/80 solution.}
    \label{fig:IL-Tdep}
\end{figure}

\pagebreak
\subsection{Complement on the SANS analysis of higher acidic concentration AcABS}

The attractive (sticky) potential is modeled as a square well potential, which depth and width are represented in the figure \ref{fig:sticky-pot_middle-concentration}.
The spherical form factor alongside the sticky hardsphere structure factor fits nicely all the data obtained for the lowest acid concentrations and those at room temperature for all the acid concentrations. The polydispersity or the sphere radius (form factor) is also taken into account.
The data measured on the intermediate acid concentrations could also be fitted with the same model, although the agreement was slightly degraded at low Q for the highest temperature. The stickiness of the potential, showing similar behavior as the lowest concentration, at least up to the phase separation, is represented in the figure \ref{fig:sticky-pot_middle-concentration}. An increase of the attractive potential between the micelles (which results from the sum of the Van Der Waals and the electrostatic repulsion) leads to the micelles flocculation. The depth of this potential increases with the temperature, in good agreement with the attractive potential interpretation that leads to the flocculation observed at higher temperature. A higher concentration of acid leads to an increase of the potential, that is attributed to the screening of the electrostatic repulsion and not to an increase of the Van der Waals forces, since the size of the micelles remains constant.     

\begin{figure}
    \centering
    \includegraphics[scale=0.5]{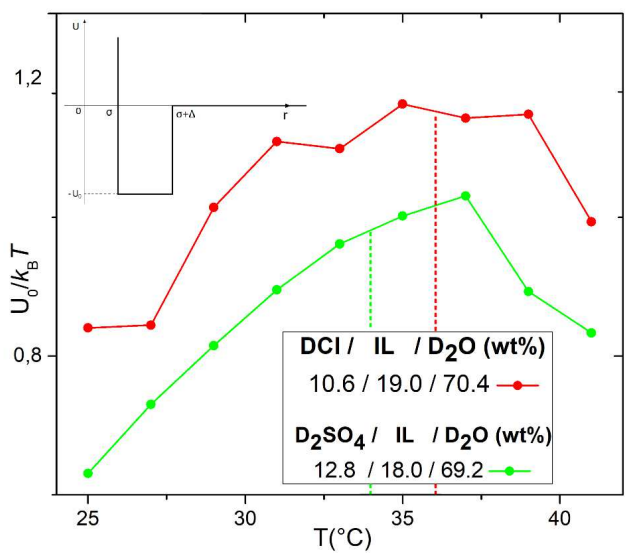}
    \caption{Temperature evolution of the stickiness of the potential for the intermediate acidic concentrations. The data are fitted with the same model for all samples : a hard sphere form factor (radius around 18 $\AA$) and sticky hard spheres structure factor. The dotted lines indicate the transition temperature.}
    \label{fig:sticky-pot_middle-concentration}
\end{figure}

\begin{figure}
    \centering
    \includegraphics[scale=0.5]{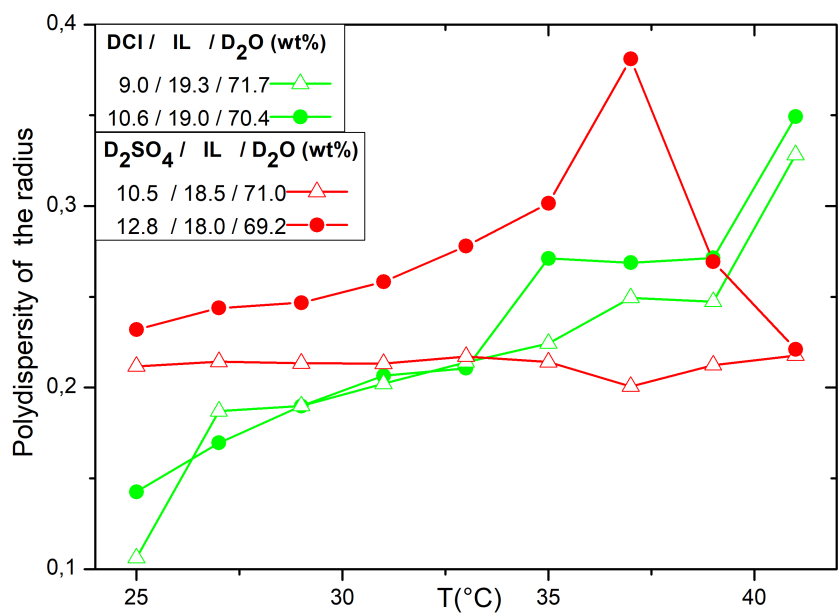}
    \caption{Polydispersity of the sphere form factor radius as a function of temperature for the lower acid concentration samples.}
    \label{fig:polydisersity-ABS}
\end{figure}

We did not find a unified model to fit all the samples we have characterized. Upon increasing the acid concentration, we got very good fitting agreement using models with coexisting shapes (spheres and cylinders for example) and bicontinuous phases. The intensity plateau expected at very low q was not reached at higher temperature, indicating the presence of very long objects or structural effects on a very large scale. We present here the results obtained for both acids with a cylindrical form factor. The radius of the cylinders was found to be roughly equal to the radius of the spherical micelles at lower concentration, and their length is given in figure 12.

\begin{figure}
    \centering
    \includegraphics[scale=0.4]{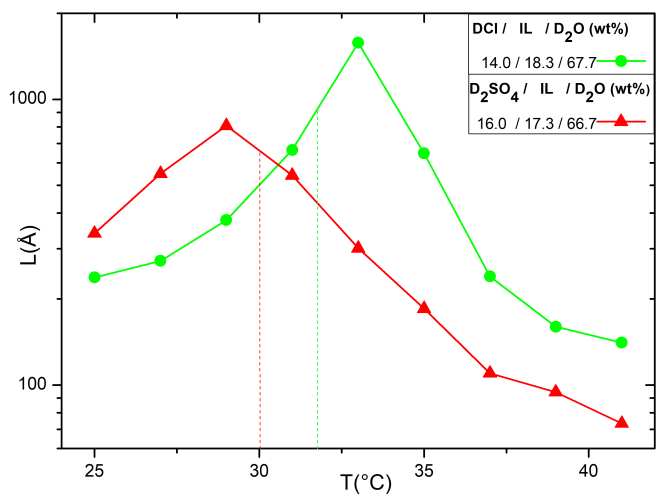}
    \caption{Length of cylinders found in the form factor for the two highest concentration samples as a function of temperature. The dotted lines indicate the transition temperature.}
    \label{fig:polydisersity-ABS}
\end{figure}

\pagebreak
\subsection{Osmotic compressibility extracted from SANS analysis of IL/water and ABS solutions}

The osmotic compressibility is the limit of the structure factor at Q=0. In the case of non interacting particles, S(0)=1. It is larger than 1 for attractive interactions, and smaller than 1 for repulsive interactions.
In order to get this value, the model fitting the experimental data was extrapolated at Q=0.

\begin{figure}
    \centering
    \includegraphics[scale=0.4]{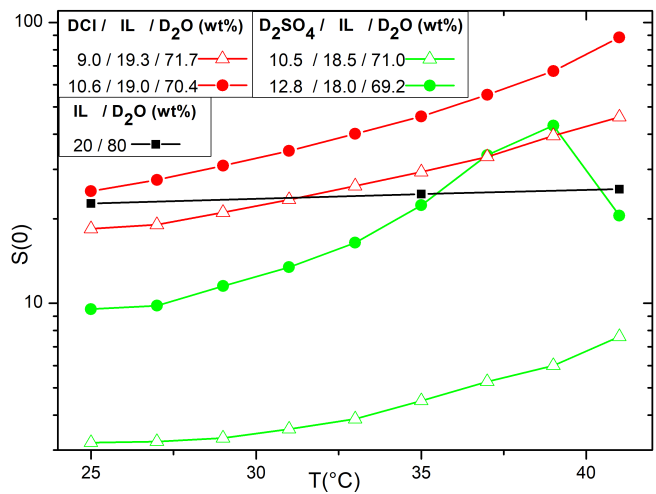}
    \caption{S(0), or osmotic compressibility as a function of temperature for binary 20$\%$ IL in water solution and two ABS solutions in DCl}
    \label{fig:osmotic-compressiblity}
\end{figure}

\end{suppinfo}


\end{document}